\documentclass[final,authoryear,1p,times]{elsarticle-ms}

\journal{To appear in Ecological Modelling, accepted 24 Oct 2016
  \hfill DOI: 10.1016/j.ecolmodel.2016.10.020}

\usepackage{url,amsmath,bm,graphicx,amssymb}

\newcommand{\captionfont}{\sf} %{\small} % or {\sf} or {}
\newcommand{\fig}[3][]{\leavevmode\centering%
\includegraphics[width=#1\textwidth]{#2.pdf}\caption{\captionfont #3}\label{fig:#2}}

\newcommand{\tab}[4][\small]{\centering%
\caption{#3}\label{tab:#2}\medskip#1%
\begin{tabular}#4\end{tabular}%
}

\newcommand{\der}[2]{\frac{{\mathrm d}#1}{{\mathrm d}#2}}
\newcommand{\dr}[2]{{\mathrm d}#1/{\mathrm d}#2}

\usepackage{hyperref}
\hypersetup{pdfstartview={FitH -32768},pdfborder={0 0 0},bookmarksopen,breaklinks}

\begin{document}

\begin{frontmatter}

\title{
Cohort aggregation modelling for complex forest stands: Spruce-aspen mixtures in British Columbia
}

\author{Oscar Garc\'{\i}a}

\address{Dasometrics, Conc\'on, Chile\fnref{f1}}

\fntext[f1]{
\copyright 2016. This manuscript version is made available under the CC-BY-NC-ND 4.0 license:\\\hspace*{5mm} \url{http://creativecommons.org/licenses/by-nc-nd/4.0/}. \ead{garcia@dasometrics.net}}

\begin{abstract}
Mixed-species growth models are needed as a synthesis of ecological knowledge and for guiding forest management. Individual-tree models have been commonly used, but the difficulties of reliably scaling from the individual to the stand level are often underestimated. Emergent properties and statistical issues limit their effectiveness. A more holistic modelling of aggregates at the whole stand level is a potentially attractive alternative.
This work explores methodology for developing biologically consistent dynamic mixture models where the state is described by aggregate stand-level variables for species or age/size cohorts. The methods are demonstrated and tested with a two-cohort model for spruce-aspen mixtures named SAM. The models combine single-species submodels and submodels for resource partitioning among the cohorts. The partitioning allows for differences in competitive strength among species and size classes, and for complementarity effects. Height growth reduction in suppressed cohorts is also modelled.
SAM fits well the available data, and exhibits behaviors consistent with current ecological knowledge. The general framework can be applied to any number of cohorts, and should be useful as a basis for modelling other mixed-species or uneven-aged stands.
\end{abstract}


\end{frontmatter}

\section{Introduction} \label{sec:intro}

There is increasing interest in managing existing forest stands with complex structure, comprised of several species and/or age classes, as well as in establishing man-made forests with those characteristics \citep{groot04a,lilles14,pretzsch15a,kabzems15,rio16}. Much progress has been made in the design and analysis of experiments or observational studies to predict and understand the development of these heterogeneous stands, but results are often inconclusive and difficult to interpret \citep{forrester15}. Appropriate mathematical models are needed to synthesize available knowledge and to reliably describe stand dynamics under various growing conditions and silvicultural alternatives.

The conceptually simplest modelling approach for complex stands is at the individual-tree level, representing tree growth and neighborhood interactions that are then aggregated to predict consequences at the stand level. Such individual-based models have a long history in forestry \citep{reventlow79,staebler51,newnham64}, and more recently have become widely used in plant ecology and other fields \citep{grimm}. Individual-tree modelling is by far the prevalent paradigm for complex stands, both for management and for research \citep{weiskittel11a,pretzsch09}. It appears, however, that forest stands can act like \emph{complex systems}, ``made up of a large number of parts that
interact in a nonsimple way'' \citep{simon62,Sayama}. These systems have (\emph{emergent}) properties at a macroscopic scale (e.~g., at the stand level) that are hard to explain simply from microscopic properties (e.~g., tree level). Several examples of such difficulties in individual-tree modelling have been identified \citep{lee16}: (a) Current models ignore positive spatial tree size correlations induced by heterogeneity in soil nutrients and moisture content, which frequently override the negative correlations expected from neighborhood competition. Either way, the analogy of tree size distributions to probability distributions based on statistical independence is questionable; size distributions are either ill-defined or plot-size dependent, and spatial correlation can introduce serious biases in predicted stand-level quantities \citep{mitchell-olds87,sambakhe14}.  (b) Plasticity allows trees to dynamically adapt their  foliage and root distribution according to space availability, causing models based on rigidly fixed tree shapes, locations and allometries  to be unrealistic \citep{strigul08}. (c) Past competitive interactions
are integrated in current tree size \citep{perry85}, and having essentially the same size variable on both sides of tree growth equations is a source of  statistical confounding that inflates fit statistics, and causes predictions to be unreliable outside the growing conditions represented in the data. (e) Growth and mortality may be unpredictable at the individual level due to sensitivity to initial conditions that are not precisely known \citep[chaos;][]{models}.
It follows that a forest stand is more than a sum of individual trees, and simple bottom-up modelling of tree behavior may fail to adequately reproduce behavior at the stand scale. This can be especially troublesome when growing conditions change over time through disturbances or a varying environment \citep[e.~g.,][]{russell15}. 

Complex or chaotic systems can become more predictable if described on a coarser scale through aggregation \citep{salas05,models}. Therefore, whole-stand modelling, with stand-level means or totals as state variables, could be an attractive alternative to individual-tree approaches in forest management. Aggregate  values like mean heights, basal area and trees per hectare can be reliably measured and are often adequate for decision-making. If necessary, size distributions may be approximated by top-down disaggregation. Despite the practical limitations of individual-based methods in their current form, they still are valuable research tools that can shed light into the issues mentioned in the previous paragraph.

Whole-stand modelling is well developed for even-aged monocultures \citep[][Section 4.2]{weiskittel11}, but it has rarely been used for managing complex stands. A notable exception is the work of Moser and associates, who produced growth and yield models projecting aggregates by species or age/size class cohorts \citep{moser69,lynch86,atta00}. Moser's models are empirical, entirely data-driven. More mechanistic biologically consistent approaches are desirable for reduced data requirements and for plausible extrapolation.
This work aims to explore cohort aggregation methodology through the development of a growth and yield model for spruce-aspen mixed stands named SAM (for Spruce-Aspen Mixtures). The new model takes advantage of existing single-species models for spruce \citep[Scube;][]{scube} and aspen \citep[TAG;][]{aspen}, extending them to represent the dynamics of two interacting cohorts differing in species and density, and possibly in size and age. The principles can be generalized to any number of cohorts. SAM is viewed as experimental, given the current gaps in knowledge and information about the dynamics of these forests. Research is still needed to fill those gaps, and possibly improve equations and statistical methods.

Following from the single-species formulations, SAM describes the state of the two-cohort mixture by each cohort's mean height, number of trees per hectare, basal area, and a measure of canopy closure.  Modelling is guided by the traditional concept of \emph{growing space} as determining \emph{resource capture} \citep{monteith94}, an abstraction that drives tree growth and mortality. The single-species models are re-written in terms of tree-level means, and stand density is expressed through its reciprocal, the mean growing space per tree. Growing space partitioning in the mixture depends on stand density and species composition, following on the classical  work of \citet{wit60}. In this instance, de Wit's results were extended to take into account differences in tree height. In addition, I allowed for possible complementarity effects that can increase or decrease resource availability in  mixtures compared to monocultures \citep{larocque13,forrester14}. These basic ideas of an equivalence between a mixture and two separate monocultures are illustrated in Fig.~\ref{fig:Figure1}. Other issues that the modelling must resolve are the relationship between cohort mean heights and the top heights used in the single-species models, and the height growth suppression observed in dominated cohorts. 

\begin{figure}[htbp]
\fig{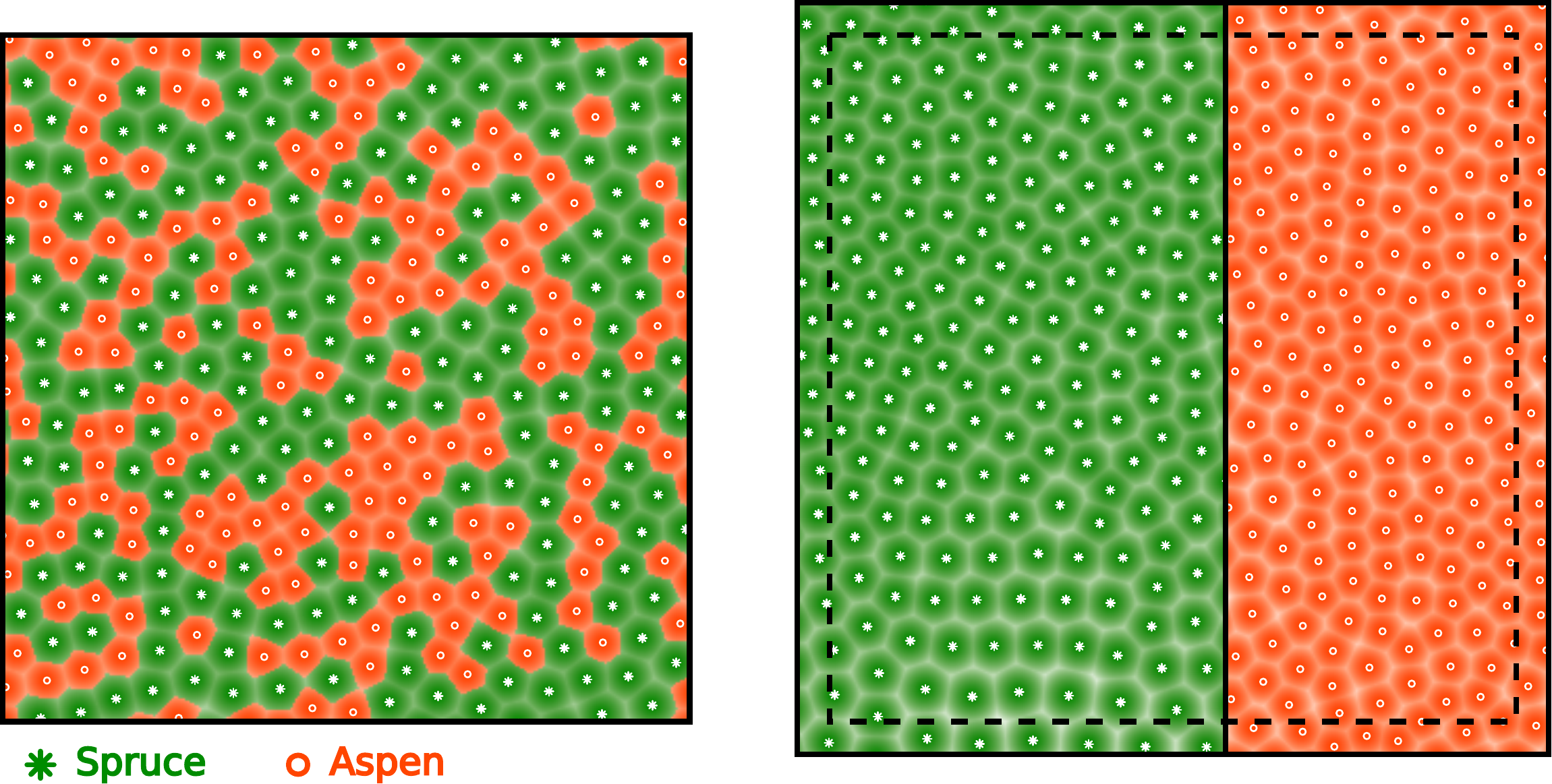}{Growing space in a mixture (left), and single-species equivalent (right). The figure assumes equal number of trees and height in both cohorts. Growing space or resource capture per tree is species-dependent. The difference in total area between mixture and monocultures reflects a difference in efficiency due to complementarity effects. Variation with tree size is introduced later in Section \ref{sec:size}. The figure is based on the fitted model parameters.}
\end{figure}

The next section contains a description of the data and of the single-species models, followed by the growing space partitioning approach.  Then top to mean height relationships, height growth suppression, and parameter estimation methods are discussed. \emph{Results} include the parameter values, residuals, and implementation details, as well as examples of model behavior. The article ends with discussion and conclusions. Appendices contain additional details and mathematical derivations.

\section{Materials and Methods}
\label{sec:methods}

    \subsection{Data}
    \label{sec:data}

Data was provided by the Forest Analysis and Inventory Branch of the British Columbia Ministry of Forests, Lands and Natural Resource Operations. Initial screening selected all the permanent sample plots from the Province that had re-measurements, and had at least 80\% by basal area of either spruce (white spruce, \emph{Picea glauca} [Moench] Voss, or interior spruce, a \emph{Picea glauca $\times$ engelmannii} hybrid complex) or trembling aspen (\emph{Populus tremuloides} Michx.).

Trees have been measured only above diameter thresholds that vary between 4 and 9 cm. To avoid complications arising from these different measurement standards and the presence of ingrowth, the numbers and diameters of the missing trees were estimated by the procedure described in Section 2.1.3 of \citet{aspen}. Only measurements where the estimated basal area of those imputed trees was less than 10\% of the total were kept, thus limiting the possible impact of imputation errors. In addition, plots with areas less than 0.04 ha, with less than 600 total trees per hectare,  or containing trees from previous generations (\emph{veterans}), were excluded.

Heights usually were only measured on a subsample of trees, and estimated by the Ministry standard height-diameter multi-year regression procedure of \citet{flewelling94}. Tree total stem volume was also computed by standard Ministry methods. Live tree data was summarized by species at the plot level to calculate mean height, trees per hectare, basal area, volume, etc.. Top heights, meaningful for largely single-species stands, were calculated in plots of $A$ hectares as the mean height of the $160 A - 0.6$ largest-diameter trees, interpolating as necessary \citep{topht,aspen}.

Growth trends and mortality were examined graphically by species and biogeoclimatic zones \citep{bec}. There was substantial representation from the SBS, BWBS, ESSF, ICH, and MS zones. Potential zone effects are interesting because the single-species spruce model Scube was developed with data from the SBS biogeoclimatic zone, while the aspen model TAG used data sources spread over all of western Canada. There were no obvious regional differences in growth patterns within British Columbia, although data variability was high and they cannot be ruled out.
Residuals from TAG predictions suggested some under-prediction of basal areas for equivalent heights, and a localization adjustment of that model for British Columbia is described later.

There was no indication in the data of the breakup that is often observed in old-growth aspen stands \citep{frey04,aspen}. However, similar catastrophic mortality was found in mature spruce, mostly for mean heights above 20--25 m, and apparently unrelated to stand composition (Fig.~\ref{fig:Figure2}). This spruce breakup does not seem to have been reported in the literature, and merits further research. Its absence in the data from the SBS zone, on which Scube is based, might be explained by a scarcity of tall trees. To avoid dealing with these unpredictable events,  measurements with spruce mean height greater than 25 m were omitted. Therefore, the model is not intended to reflect old-growth disintegration or successional processes.

\begin{figure}[htbp]
\fig{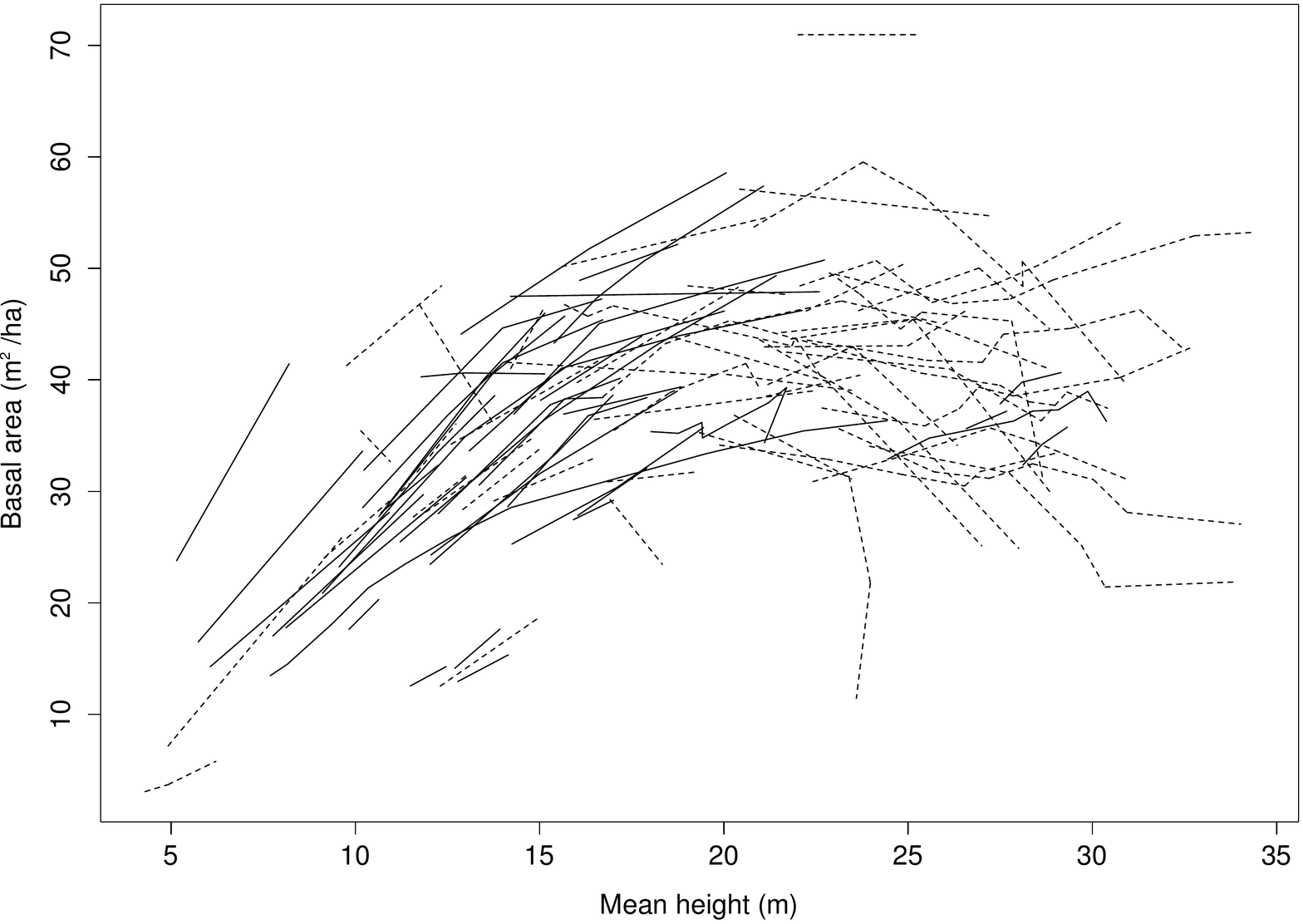}{Observed trends of spruce basal area \emph{vs} height for spruce-dominated plots ($>$ 70\% by basal area), indicating heavy  mortality in tall stands. Continuous lines are for the SBS and dashed lines are for other biogeoclimatic zones.}
\end{figure}

Most measurements have a substantial component of species other than spruce or aspen. Since this is a spruce-aspen model, only measurements with less than 10\% of other species by basal area were retained. This threshold was chosen as a compromise between adhering to the two-cohort concept and maintaining an acceptable sample size. Closeness to the two-cohort ideal was enhanced through adjusting plot areas multiplying by the proportion of spruce or aspen, i.~e., dividing the previously calculated basal areas and trees per hectare by this quantity. The final simulation software has the option of reversing this adjustment to represent stands with a moderate proportion of other species. Given the exclusion of multiple species, SAM may be more representative of man-made mixtures that of typical natural stands.

Successive plot measurements were paired for estimating the model parameters. After eliminating some clearly anomalous observations, 118 interval pairs were available, of which 13 had no aspen, 23 had no spruce, and 82 contained both species in various proportions. Tables \ref{tab:data1} and \ref{tab:data2} give descriptive pair statistics.

\begin{table}[htbp]
  \tab[\scriptsize]{data1}{Growth interval characteristics. 118 measurement pairs from 88 plots.}{
      {lccccccc}
      \hline
      &   Mean  & SD  & Minimum & Quartile 1 & Median  & Quartile 3 & Maximum \\
      \hline
      Plot size (ha)  &  0.0772 &  0.0232 & 0.0400 & 0.0506 & 0.0809 & 0.1012 & 0.1200 \\
      Initial year & 1986 & 8.48 & 1949  &  1981 &   1987 &  1992 &   1997 \\
      Interval length (years) & 10.58 & 2.07 & 5.00  & 10.00  & 10.00  &  11.00 &  18.00 \\
      Spruce \% basal area  &  45.22 & 37.48 & 0.00  & 8.25 & 39.51 &  81.01 & 100.00 \\
      Mean height difference, &&&&&&&\\
       aspen minus spruce (m)  & 5.81 & 4.82 & -2.57  & 2.79 &  4.85 &  8.55 & 20.19 \\
      \hline
  }
\end{table}

\begin{table}[htbp]
  \tab{data2}{Data statistics. 118 measurement pairs from 88 plots.}{
  {lcccccc}
      \hline
      & \multicolumn{2}{c}{Spruce} & \multicolumn{2}{c}{Aspen} & \multicolumn{2}{c}{Total} \\
      &   Mean  & SD  &  Mean  & SD  &   Mean  & SD  \\
      \hline
       \multicolumn{7}{c}{Measurement pair means} \\
      Mean height (m)  & 13.97 & 4.00 & 18.70 & 4.55 & 16.45 & 4.90 \\
      Trees / ha  & 785.4 & 645.9 & 1039.3 & 1237.1 & 1824.8 & 1067.6 \\ 
      Basal area (m$^2$/ha) & 18.49 & 16.79 & 21.26 & 14.95 & 39.76 & 10.22 \\
      Arith.~mean dbh (cm)  & 15.76 & 4.46 & 18.68 & 5.74 & 17.29 & 5.36 \\
       \multicolumn{7}{c}{Annual increments} \\
      Mean height (cm/yr)  & 2.703 & 0.845 & 2.345 & 1.049 & 2.403 & 0.785 \\
      Trees / ha (trees/ha--yr) & -8.39 & 14.28 & -20.86 & 28.80 & -29.25 & 29.99 \\
      Basal area (m$^2$/ha--yr) & 0.3493 & 0.3956 & 0.1850 & 0.3191 & 0.5343 & 0.4144 \\
      Arith.~mean dbh (mm/yr)  & 2.779 & 1.076 & 2.680 & 1.251 & 2.575 & 0.860 \\
      \hline
  }
\end{table}

Data for some auxiliary relationships described in the appendices are summarized there.

    \subsection{Single-species models}
    \label{sec:mono}

Two existing dynamic growth and yield models were used as a starting point, the Scube model for spruce-dominated stands \citep{scube}, and the TAG model for aspen-dominated stands \citep{aspen}. In this section the relevant equations are shown, and are then re-written in a form suitable for application to the mixture cohorts.

Both Scube and TAG have a similar structure, describing a stand through 4 state variables: top height $H$ (m), stand density $N$ (trees/ha), a volume or biomass proxy $W = BH$ (m$^3$/ha) where $B$ is basal area, and a measure of relative site occupancy $\Omega$ (dimensionless, not directly observed). The state evolution is determined by a system of differential equations
\begin{subequations}  \label{eq:odes}
\begin{align}
    \der{H}{t} &= (p_2/p_3) H [(p_1 / H)^{p_3} - 1]	  \label{eq:ode1}	\\
	\der{N}{H} &= - p_4 H^{p_5} N^{p_6}		\label{eq:ode2}	\\
	\der{W}{H} &= p_7 \Omega H N^{p_8} + p_{9} \frac{W}{N} \der{N}{H}	\label{eq:ode3} \\
	\der{\Omega}{H} &= p_{10} H (1 - \Omega)  \label{eq:ode4}\;,
\end{align}
\end{subequations}
where $p_1, \dots, p_{10}$ are parameters shown in Table~\ref{tab:pars}. The original equations also include a species composition variable, but only the form for pure stands is needed here. Integration produces transition functions that predict state variable values at any future time given their current values.

\begin{table}[htbp]
  \begin{minipage}{\textwidth}
  \tab{pars}{Single-species parameter values}{
    {lcc}
        \hline
        Parameter & Spruce  & Aspen\footnote{
        The aspen parameters are those in the current version of TAG (\url{http://forestgrowth.unbc.ca/tag}), with some differing from the published ones.
        } \\
        \hline
        $p_1$    & $283.87 q^{0.5137}$  & $8.568 + 1342 q$  \\
        $p_2$    & $q$  & $q$  \\
        $p_3$    & 0.5829  & 1  \\
        $p_4$    & $4.5759\times 10^{-15}$ & 0.0004561  \\
        $p_5$    & 5.009  & 0  \\
        $p_6$    & 2.9895 & 1.6159  \\
        $p_7$    & 0.5345 & 0.8560  \\
        $p_8$    & 0.3 & 0.2029  \\
        $p_9$    & 0.3  & 0.5  \\
        $p_{10}$ & $2.2\times 0.023474$ & 0.03  \\
        $p_{11}$ & $1.815\times 10^{-6}$ & $1 / 15605$  \\
        $p_{12}$ & 2.2 & 2.6  \\
        \hline
    }
    \end{minipage}
\end{table}

Eq.~\eqref{eq:ode1} is a self-contained height growth and site index sub-model, in which $p_1$ and $p_2$ are functions of a site quality parameter $q$ (Table~\ref{tab:pars}).
The value of $q$ may be derived from conventional site index or from biogeoclimatic classification or other methods. In principle, $q$ could vary over time following climate change, but it has been kept constant in the implementations.
Mortality, eq.~\eqref{eq:ode2}, and wood accumulation, eq.~\eqref{eq:ode3}, depend on stand height and density. The occupancy factor $\Omega$ reduces growth when the site potential is not fully utilized, e.~g., in young open-canopy stands or following thinning. $\Omega$ is initialized depending on initial stand density, and tends to 1 according to eq.~\eqref{eq:ode4}. Occupancy can be abruptly reduced by thinning.

These models are somewhat atypical in not including basal area or diameter as drivers on the right-hand sides. This is consistent with physiological knowledge, the amount of wood deposited on the stems should not have a significant causal effect on growth or mortality \citep{scube}. In addition, the presence of volume or basal area on both sides of a growth equation could cause statistical confounding and extrapolation issues similar to those found in individual-based models \citep{lee16}.  

In preparation to applying the models to mixture components, we express  equations \eqref{eq:ode2} and \eqref{eq:ode3} at the level of individual trees, predicting a mean death probability or mortality fraction $- (\dr{N}{H}) / N = - \dr{\ln N}{H}$, and the mean $\bar W = W / N$. Using $\dr{W}{H} = N \dr{\bar W}{H} + \bar W \dr{N}{H}$,
\begin{align*}
    \der{\ln N}{H} &= - p_4 H^{p_5} N^{p_6 - 1}	\\
	\der{\bar W}{H} &= p_7 \Omega H N^{p_8 - 1} + (p_{9} - 1) \bar W \der{\ln N}{H}
\end{align*}
Also, instead of $N$ on the right-hand sides, a biologically more meaningful predictor is the average growing space per tree $\bar S = 1 / N$, preserving the area units to simplify the presentation ($\bar S$ in hectares per tree, although m$^2$ would be more practical). With these substitutions, equations \eqref{eq:ode2} and \eqref{eq:ode3} become
\begin{subequations}  \label{eq:mix}
\begin{align}
    \der{\ln N}{H} &= - p_4 H^{p_5} \bar S^{1 - p_6}	\label{eq:mix1} \\
	\der{\bar W}{H} &= p_7 \Omega H \bar S^{1 - p_8} + (p_{9} - 1) \bar W \der{\ln N}{H} \;. \label{eq:mix2}
\end{align}
\end{subequations}
It is assumed that in this form the models are applicable to the respective species cohort.

Derivatives with respect to $t$ for each cohort can be obtained multiplying equations \eqref{eq:mix} by $\dr{H}{t}$ from eq.~\eqref{eq:ode1}. The system of differential equations can then be integrated numerically, given the dependency of the growing space $\bar S$ on the other state variables.

For a dominated cohort it was found necessary to multiply eq.~\eqref{eq:ode1} by a height- and density-dependent height growth suppression factor discussed in Section \ref{sec:hsuppr}. Therefore, the analytical solution of eq.~\eqref{eq:ode1} from the single-species models cannot be used, and the height growth equation  has to be included in the numerical integration.

The occupancy factor $\Omega$ can still be calculated from an analytical integral of eq.~\eqref{eq:ode4}:
\begin{equation}
    \Omega = 1 - (1 - \Omega_0) \exp[-p_{10} (H^2 - H_0^2)/2] \;.  \label{eq:omega}
\end{equation}
In unmanaged stands $\Omega$ can be initialized at breast-height ($H_0 = 1.3$ m) with
\begin{equation} \label{eq:omega0}
    \Omega_0 = 1 - (1 - \min\{p_{11} / \bar S_0, 1\})^{p_{12}} \;,
\end{equation}
where $\bar S_0$ may be estimated back-projecting mortality from the first plot measurement \citep{scube,aspen}. Most of the development of these stands occurs at near canopy closure ($\Omega \approx 1$), so that the  impact of these occupancy equations is usually small.

As mentioned before, TAG had to be localized to better fit the data from British Columbia. This was done by multiplying the right-hand sides of \eqref{eq:mix} by a localization parameter $\lambda$, thus adjusting the mortality rate and the volume or biomass growth relative to height growth.

Some computational aspects are described later in the text. Full details with computer code are contained in the Supplementary Material.

The mixture growing-space relationships are discussed in the next section. Top height is a convenient variable in single-species stands, but its definition and measurement is problematic in mixtures. Therefore, top height is approximated by a function of mean height and growing space (Section \ref{sec:topmean}). The growth model specification is completed by the height suppression relationships in Section \ref{sec:hsuppr}.

    \subsection{Growing space}
    \label{sec:S}

Stand density  and composition affects competition intensity, with individual growth and mortality depending on the space available to each plant (eq.~\eqref{eq:mix}).
In an even-aged monoculture the mean growing space per plant is simply $\bar S = 1 / N$. The situation is more complex for species or age/size cohorts, and finding ``the most appropriate ways to estimate the species proportions by area in mixed stands'' \citep{dirnberger14} has recently received considerable attention in the literature \citep[e.~g.,][]{sterba14,pretzsch15a,rio16}. We look for a simple approach appropriate to dynamic modelling and consistent with biological principles.

The next section gives a compact derivation of the relevant results from \citet{wit60}. These are extended in Sect.~\ref{sec:compl} to represent possible complementarity effects. Section \ref{sec:size} generalizes the model further to deal with the consequences of different tree sizes.

        \subsubsection{Interspecific partitioning}
        \label{sec:ic}

The simplest way of apportioning area would be uniformly, ignoring any species or size differences. With the subscripts $s$ and $a$ denoting spruce and aspen, respectively,
\begin{equation} \label{eq:unif}
    \bar S_s = \bar S_a = \frac{1}{N_s + N_a} \;.
\end{equation}
Then, the mean individual-tree growth and mortality rate for the spruce component would equal those of a monoculture with a density of $N_s + N_a$ trees/ha, and analogously for aspen. The fraction of the total space captured by spruce, $S_s = N_s \bar S_s$, equals its proportion in the mixture, $P_s = N_s / (N_s + N_a)$.
A similar approach is used in the extension of TIPSY to mixed-species stands \citep{tipsy}.
    
More generally, one may expect that the species differ in competitive strength, with one of them intruding into the space of neighbors belonging to the other species. Instead of $\bar S_s = \bar S_a$, assume that
\begin{equation} \label{eq:crowding}
    \bar S_s = k \bar S_a \;,
\end{equation}
 where $k$ is a \emph{relative crowding coefficient} of spruce with respect to aspen \citep[][p.~15]{wit60}. Solving this equation together with the total growing space constraint
 \begin{equation} \label{eq:noc}
    S_s + S_a = N_s \bar S_s + N_a \bar S_a = 1 \;,
    \end{equation}
it is found that the mean growing spaces are
\begin{equation} \label{eq:dewit}
    \bar S_s = \frac{k}{k N_s + N_a} \;, \quad
    \bar S_a = \frac{1}{k N_s + N_a} \;.
\end{equation}
It can be verified that this agrees with de Wit's equations, which are expressed in relative terms:
\[
    S_s = \frac{k P_s}{1 + (k - 1) P_s} \;, \quad
    S_a = \frac{k^{-1} P_a}{1  + (k^{-1} - 1) P_a} \;.
\]

        \subsubsection{Complementarity}
        \label{sec:compl}

The above assumes a fixed amount of resources per unit area shared between the two cohorts. However, deviations are possible in species mixtures. For instance, different rooting depths can make more soil resources available \citep{kelty92}. Various facilitation and other mechanisms with similar effects can be active and difficult to separate, and are often collectively described as \emph{complementarity} \citep{forrester14}. We model these by increasing (or decreasing) the growing space or resource availability represented on the right-hand side or eq.~\eqref{eq:noc}.

A simple modification of eq.~\eqref{eq:noc} might add a term proportional to $4 P_s P_a$ to the right-hand side. This is 0 for a single-species stand, and reaches a maximum value of 1 when $P_s = P_a = 1/2$. A slightly more complex form that reflects the species relative crowding was used:
\begin{equation} \label{eq:cterm}
    4 \left(\frac{k P_s}{k P_s + P_a} \right)
    \left(\frac{P_a}{k P_s + P_a} \right)
     = \frac{4 k N_s N_a}{(k N_s + N_a)^2} \;.
\end{equation}
The maximum is now at $k P_s = P_a$ or $k N_s = N_a$, which corresponds to equal growing spaces: $S_s = N_s \bar S_s = N_s k \bar S_a = N_a \bar S_a = S_a$.  Eq.~\eqref{eq:noc} becomes 
\begin{equation} \label{eq:compl}
    N_s \bar S_s + N_a \bar S_a = R \;,
\end{equation}
with
\begin{equation} \label{eq:R}
    R = 1 + c \frac{4 k N_s N_a}{(k N_s + N_A)^2} \;,
\end{equation}
$c$ being a \emph{complementarity parameter}. A power of eq.~\eqref{eq:cterm} with the exponent as an additional shape parameter could provide additional flexibility, but would likely lead to over-parametrization with our data.

Solving the linear system formed by equations \eqref{eq:crowding} and \eqref{eq:compl} gives a generalization of deWit's eq.~\eqref{eq:dewit}, where the right-hand sides are multiplied by $R$ from eq.~\eqref{eq:R}.
Complementarity can be assessed by testing the hypothesis $c = 0$.

        \subsubsection{Size effects}
        \label{sec:size}

As noted by \citet[][Ch.~8]{wit60}, the crowding coefficient can be sufficient for characterizing final yields, but at intermediate times the model is strictly valid only if the growth curves are proportional \citep[see also][]{wit65,firbank85}. Cohort size differences become important if these vary substantially over time, as can be expected from the dynamics of competition. Regeneration lags can accentuate size diferences \citep{kabzems04}.

A simple mechanistic individual-based model was used to derive a plausible functional form for the effects of size. It is based on two ideas:
\begin{enumerate}
  \item Trees exert a competitive pressure on available resources depending on species, size, and distance. Assume that the pressure (or \emph{influence function}) is proportional to the height of a paraboloid of revolution anchored at the top of the tree, and that at each point in the plane the resource is captured by the tree with the highest influence (Fig.~\ref{fig:Figure3}). The TASS model of \citet{mitchell75} is based on a similar concept, viewing the influence profiles as physical crowns subject to mechanical neighbor interference. \citet{gates79} found, given some plausible premises, that the crown profile should be a power function, and that the parabolic case produces polygonal tessellations like those often used in forest growth models (Fig.~\ref{fig:Figure3}a). The profile can be interpreted more generally as competitive strength, or as a shading potential that can extend beyond the crown surface \citep{garcia14}. One can relax slightly the assumption by allowing deformations due to neighborhood interactions, provided that the profile cross-sectional area for a tree of height $H_i$ at a level $z \leq H_i$ is proportional to the distance from the top, $b_i (H_i - z)$, where $b_i$ is a species-dependent proportionality constant.
  
  \item The \emph{perfect plasticity approximation} (PPA) postulates that tree crowns are displaced horizontally by stem leaning and by differential branch growth so as to balance competitive pressure on all sides \citep{strigul08,plasticity}
\end{enumerate}
These two assumptions imply that the individual growing spaces fill the plane at a certain closure level $z^*$ (Fig.~\ref{fig:Figure3}b).

\begin{figure}[htbp]
\fig{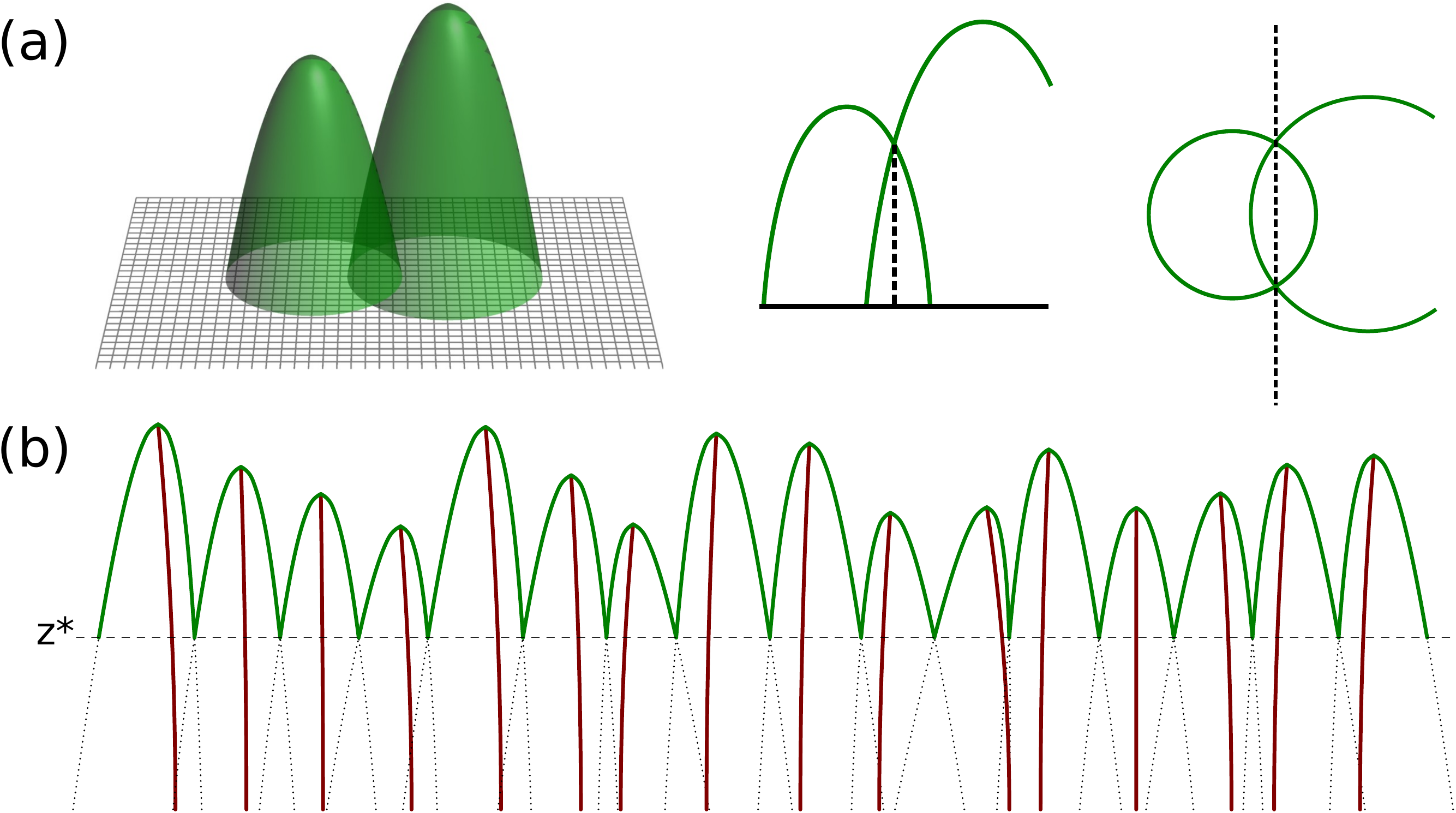}{(a) Parabolic crown profiles or influence functions intersect on a polygonal tessellation \citep[after][]{gates79}. (b) The perfect plasticity approximation assumes that inter-tree competition intensity is balanced.}
\end{figure}

Averaging over the spruce and aspen trees, the cohort mean growing space is
\begin{equation} \label{eq:SH} 
    \bar S_s = b_s (\bar H_s - z^*) \;, \quad
    \bar S_a = b_a (\bar H_a - z^*) \;.
\end{equation}
Solving the second equation for $z^*$ and substituting into the first,
\begin{equation} \label{eq:noz}
    \bar S_s = (b_s / b_a) \bar S_a + b_s (\bar H_s - \bar H_a) \;. 
\end{equation}
For the special case of equal heights this coincides with eq.~\eqref{eq:crowding}, writing $b_s / b_a = k$.
Writing also $b_a \equiv b$, the solution of the linear system formed by equations \ref{eq:noz} and \ref{eq:compl} is:
\begin{subequations}  \label{eq:spac}
\begin{align}
    \bar S_s =& \frac{k}{k N_s + N_a} \left[ 1 + b N_a (\bar H_s - \bar H_a) + c \frac{4 k N_s N_a}{(k N_s + N_a)^2} \right] \\  
    \bar S_a =& \frac{1}{k N_s + N_a} \left[ 1 + b N_s (\bar H_a - \bar H_s) + c \frac{4 k N_s N_a}{(k N_s + N_a)^2} \right] \;.
\end{align}
\end{subequations}
If the calculated space is negative it must be set to 0, indicating complete suppression by the dominant cohort.

In open-canopy stands the calculations above can result in a negative $z^*$. Setting $z^* = 0$ in those cases would make the total growing space smaller than $R$, reflecting the less than full site occupancy \citep{nance88}. It was preferred, however, to take $z^*$ as a potential closure level and to model occupancy through the $\Omega$ state variables of the single-species models.

Parabolic crown or influence profiles generate a polygonal tessellation with growing spaces linear on tree height. Some research suggests, however, that conical profiles with space proportional to $(H_i - z^*)^2$ might be more realistic \citep{plasticity,lee16}. In that case it is not possible to obtain simple exact results analogous to equations \eqref{eq:noz} and \eqref{eq:spac}, but these can still be valid as a first-order approximation. \ref{app:part} extends the results to other profiles and to any number of cohorts. Those cohorts can correspond to different species, or to age or size classes within a same species for modelling uneven-aged stands.

    \subsection{Heights}
    \label{sec:hts}

      \subsubsection{Top \emph{vs.}\ mean height}
      \label{sec:topmean}

Top height is convenient in pure even-aged stands because its development can be predicted with a simple density-independent sub-model, and because it relates to a useful measure of site quality, site index. In complex stands, however, an operational definition of top height is problematic \citep{zingg94}. It is probably best then to consider a cohort top height as a variable that is not directly observable, pertaining to a hypothetical pure stand of the same species, age, and site quality. Therefore, the top height used in the single-species models is linked here to the cohort mean height and mean growing space.

Data from essentially pure spruce and aspen stands were used to obtain relationships between top height $H$, mean height $\bar H$, and mean growing space $\bar S$. Given the small sample sizes, a much larger data set from loblolly pine plantations was also used to guide the selection of suitable equation forms. Details are shown in \ref{app:topmean}. The following equations were obtained:
\begin{subequations} \label{eq:hconv}
\begin{align}
    H_s / \bar H_s &= \exp\left[0.10439 \max\left\{-\ln(150 \bar S_s), 0\right\}^{1.159}\right] \quad \text{ (spruce),} \\ 
    H_a / \bar H_a &= \exp\left[0.04732 \max\left\{-\ln(150 \bar S_a), 0\right\}^{1.438}\right] \quad \text{ (aspen).}
\end{align}
\end{subequations}
These can be used for estimating $H$ from $\bar H$, or vice-versa.

      \subsubsection{Height growth suppression}
      \label{sec:hsuppr}

In multi-layered mixtures it is often observed that the height growth of the dominated species is reduced in relation to that expected in a monoculture \citep[e.~g.,][]{kabzems15}. More generally, a similar suppression of shorter by taller cohorts can occur in uneven-aged stands, or among trees in individual-based models. An initial model version ignoring this did not provide good predictions for some stands, and therefore a height suppression element was introduced.

Height growth suppression is commonly modelled as the product of a ``potential'' height growth rate and a ``modifier'' that depends on competitive pressure \citep[][Sec.~6.3.1]{salas08, weiskittel11}. Here the potential top height growth predicted by eq.~\eqref{eq:ode1} was reduced by a factor determined by the other species density and by the difference in top heights. These are the same variables affecting the growing space in eq.~\eqref{eq:spac}. The description that follows is for the usual case where spruce has the smallest height. The opposite situation is rare in our data (Table \ref{tab:data1}), but for completeness the analogous relationship with the cohort roles reversed was incorporated into the final model.

The spruce height growth reduction must increase with $N_a$ and with $H_a - H_s \equiv \Delta H$, being null if $N_a = 0$ or $\Delta H \le 0$. Once the spruce is below the base of the aspen canopy there should be no further reduction, that is, an asymptote for increasing $\Delta H$ would be expected. It makes sense also for the reduction to be asymptotic in $N_a$. This can be achieved with a reduction factor (modifier) of the form
\[
  1 - \alpha F_1(N_a) F_2(\Delta H) \;,
\]
where $F_1$ and $F_2$ are growth functions with origin 0 and asymptote 1 (or cumulative distribution functions of non-negative variables), and $\alpha$ is a parameter determining the maximum reduction factor $1 - \alpha$. \citet{salas08} used a related approach for individual trees.

For a reasonably simple expression one may choose $F_1$ and $F_2$ as Hossfeld IV functions:
\[
  F(x) = \frac{x^\gamma}{x^\gamma + \beta^\gamma} = 1 / [1 + (\beta / x)^\gamma]
\]
if $x \geq 0$ or 0 otherwise, with $\beta$ and $\gamma$ being positive parameters \citep{grex}. If over-parametrization due to limited data is a concern, this may be simplified further by fixing $\gamma$ at a suitable value. Experimentation showed $\gamma = 4$ to be near-optimal, producing a plausible sigmoid shape.

The spruce modifier is then
\begin{equation} \label{eq:modif}
  1 - \frac{\alpha}{[1 + (\beta_1 / N_a)^{\gamma_1}][1 + (\beta_2 / \Delta H)^{\gamma_2}]} \;,
\end{equation}
or 1 if $N_a = 0$ or $\Delta H \leq 0$. As mentioned above, the same equation form was used for the aspen height growth modifier, substituting $N_s$ for $N_a$ and $- \Delta H$ for $\Delta H$.

    \subsection{Parameter estimation}
    \label{sec:par}

The model parameters to be estimated are $k$, $b$ and $c$ from eq.~\eqref{eq:spac}, $\alpha$, $\beta_1$ and $\beta_2$ (and possibly $\gamma_1$ and $\gamma_2$) from eq.~\eqref{eq:modif}, and the localization parameter $\lambda$ (Sec.~\ref{sec:mono}). These were obtained through model projections for interval data \citep{gadow99}, specifically, from periodic annual increment (PAI) residuals over pairs of consecutive measurements. \citet{scube,aspen} compared this and other estimation strategies, and found only minor differences in the resulting model predictions. At each measurement the state of the stand was characterized by the mean height, density, and basal are per hectare for each species, i.~e., by a vector $(\bar H_s, \bar H_a, N_s, N_a, B_s, B_a)$. The stands in the database are unthinned and are expected to have reached full or nearly full canopy closure, so that for estimation the occupancy factor was fixed at $\Omega_s = \Omega_a = 1$.

The model has a hierarchical aspect, involving also site quality parameters $q$ specific to each stand and cohort. These are functionally related to conventional site indices through the integral of eq.~\eqref{eq:ode1} at some reference age. More generally, $q$ can vary over time, reflecting climate variability and the effect of pests \citep{fontes10}. An initial model version computed $q$-values for each measurement interval from the observed heights and densities, effectively eliminating the time variable by expressing basal area growth and mortality relative to height growth \citep{aspen}. Introduction of height growth suppression, however, required a more complex two-stage approach: at each iteration, $q$ was  first estimated for each plot minimizing squared height residuals, and then used in the current parameter estimation step. The spruce $q_s$ and aspen $q_a$ were constrained to satisfy
\begin{equation} \label{eq:siteconv}
  \text{Spruce Site Index} = 3.804 + 0.7978 \times \text{Aspen Site Index}
\end{equation}
\citep{nigh02a}.

Fitting was performed based on residuals of basal area and of the logarithm of stand density. The logarithm $\ln N$ was preferred to $N$ because it makes the residuals more symmetric and homoscedastic. Parameter values affect jointly basal area growth and mortality, and these variables can be expected to be correlated, so that this is a multiresponse parameter estimation problem \citep[][Ch.~11]{boxdraper,bates85,view,SeberWild}. Based on a Bayesian analysis, \citet{boxdraper} recommended minimizing the determinant of the residual sum of squares matrix. The same criterion was obtained by \citet[][Sec.~4--9]{bard} as the maximum likelihood estimator under the usual normality assumptions. As shown by \citet{SeberWild}, it is equivalent to quasi-maximum-likelihood or generalized least-squares estimation in more general distribution-free settings. Several methods of approximate computation have been proposed \citep[][Ch.~11]{bates85,SeberWild}. However, with modern computing power, direct numerical optimization of the Box-Draper criterion with general-purpose software has become feasible. Parameter estimates were calculated using the \texttt{optim} R function \citep{R} to minimize the logarithm of the determinant of the $6 \times 6$ sum of squares matrix of PAI residuals for $H_s$, $H_a$, $\ln N_s$, $\ln N_a$, $B_s$ and $B_a$.

It is interesting to test statistically for complementarity effects, that is, to test if $c$ is significantly different from 0. A likelihood ratio test suggested by \citet[][p.~538]{SeberWild} was used. It is based on the determinant of the sum of squares matrix for the full model, $|\hat V|$, and that for the model with $c=0$, $|\hat V_0|$. Under the null hypothesis $c=0$ and with a sample of $n$ observations, $n(\ln|\hat V| - \ln|\hat V_0|)$ has asymptotically a chi-squared distribution with 1 degree of freedom. The likelihood ratio was useful also for testing significance of localization and other parameters.

For the residual computations, state projections from the start to the end of a measurement interval were performed as follows. First, $q_s$ and $q_a$ values were obtained as explained above, and initial top heights $H_s, H_a$ were calculated from the mean heights and densities with equations \eqref{eq:spac} and \eqref{eq:hconv}. These heights were used to compute the initial state vector
\[ (H_s, H_a, \ln N_s, \ln N_a, \bar W_s, \bar W_a) \]
in the system of differential equations \eqref{eq:ode1}, \eqref{eq:mix1}, \eqref{eq:mix2}. The differential system was integrated numerically from the initial to the final interval time with function \emph{ode} of the \emph{deSolve} R package \citep{deSolve}. At each step, equations \eqref{eq:spac} and \eqref{eq:hconv} have to be solved-back numerically for $\bar S_s$ and $\bar S_a$. Finally, the projected mean heights, densities and basal areas $B = \bar W N / H$ are extracted from the integration output. Section \ref{sec:mod} gives a summary of the projection procedure.

Full computational details and computer code are included in the Supplementary Material.

\section{Results}
\label{sec:res}

    \subsection{Estimates}
    \label{sec:est}

A number of optimization runs were performed, with and without $c$, $\lambda$, or height growth suppression, with various values for $\gamma$, and using the Nelder and Mead or the BFGS options of \texttt{optim}. Different starting points were used to guard against local optima. Results were consistent, with typical run times, starting from rough initial estimates, of 20 to 40 minutes on a modern personal computer.

Accounting for height growth suppression was clearly necessary. Fixing $\gamma$ at 4 was better than at 2 or 3, although the differences were not statistically significant. Leaving $\gamma_s$ and $\gamma_a$ as free parameters did not produce any appreciable improvement. The final parameter estimates (with $\gamma_s = \gamma_a = 4$) are given in Table~\ref{tab:ests}, together with approximate standard errors. The standard errors are based on the observed information matrix given by the inverse Hessian of the optimized log-likelihood \citep[][Sec.~16.3]{efron78,SeberWild,venables02}.

\begin{table}[htbp]
  \tab{ests}{Parameter estimates.}{
  {lcc}
    \hline
    Parameter & Estimate & Standard error \\
    \hline
    $k$ & 1.317 & 0.0669 \\
    $b$ & $0.2153 \times 10^{-4}$ & $0.0460 \times 10^{-4}$ \\
    $c$ & 0.2078 & 0.0623 \\
    $\alpha$ & 0.2876 & 0.0316 \\
    $\beta_1$ & 33.81 & 50.5 \\
    $\beta_2$ & 3.459 & 0.609 \\
    $\lambda$ & 1.448 & 0.0483 \\
    \hline
    }
\end{table}

The higher competitive strength of spruce relative to aspen implied by the relative crowding coefficient $k = 1.317$ agrees with the similar, although higher value $6.543 / 3.663 = 1.786$ found by \citet{lee16} in spruce-hardwood mixedwoods. The complementarity parameter $c$ was significantly different from 0 ($p = 8.2 \times 10^{-11}$), giving a maximum growing-space gain of 21\% for mixtures relative to monocultures (eq.~\eqref{eq:R}). The aspen localization parameter $\lambda$ was significantly different from 1 ($p = 9.0 \times 10^{-10}$), indicating larger diameters for a common height in British Columbia compared to the predictions of the western Canada model. A similar $\lambda$ was obtained using only pure-aspen plots.

\begin{figure}[htbp]
  \fig{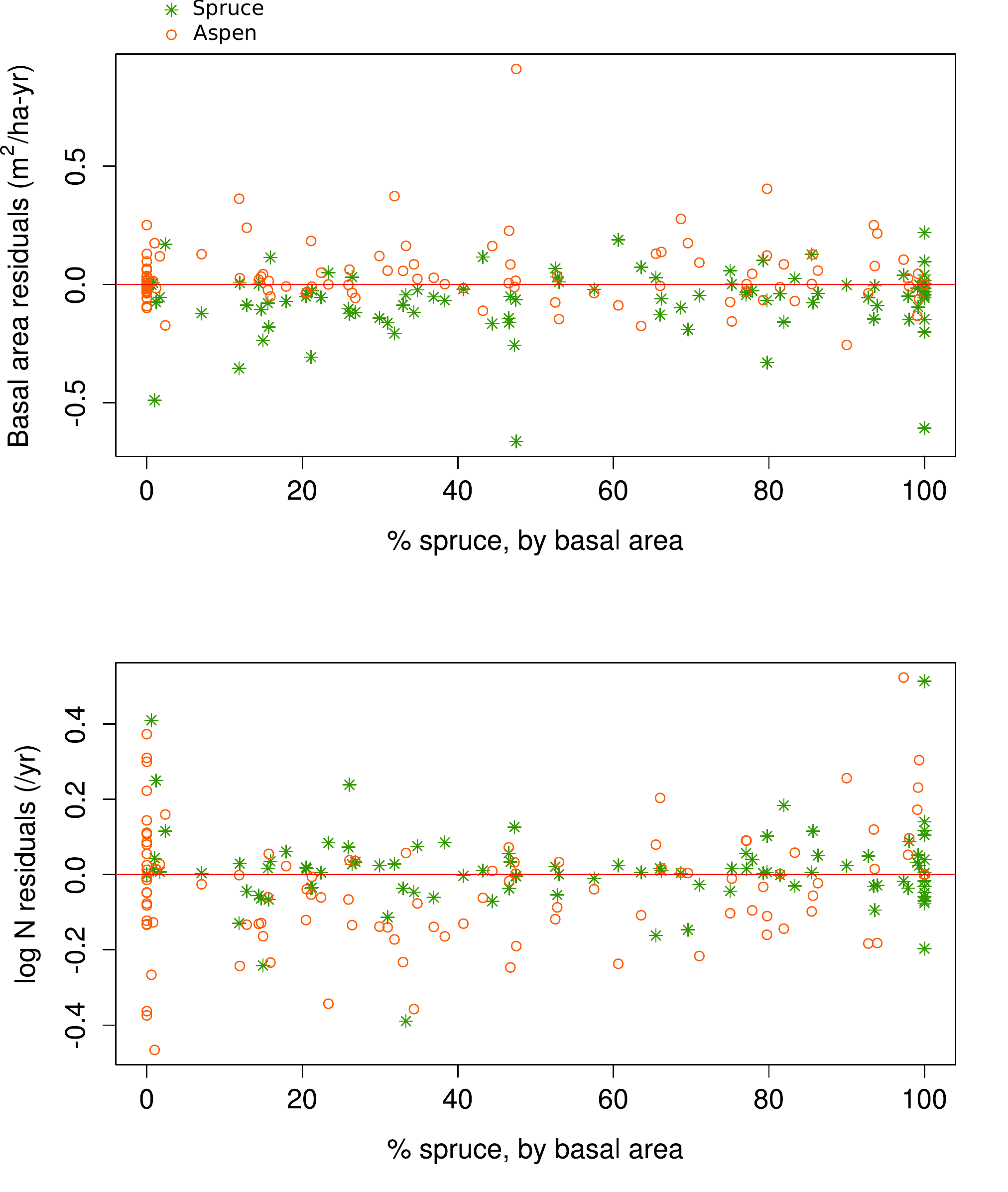}{Measurement interval periodic annual increment residuals (see text).}
\end{figure}

Figure \ref{fig:Figure4} shows the basal area and mortality residuals, related to species composition. The residual is the difference between the observed and predicted PAI, i.~e., the difference of the values at the second measurement of the interval pair divided by the interval length.  Spruce basal area percentages are ratio-of-means averages for the interval.

    \subsection{Model summary and implementation}
    \label{sec:mod}

At any point in time the stand can be described by the 8 state variables $\bar H_s, \bar H_a, N_s, N_a, B_s, B_a, \Omega_s, \Omega_a$. The growth model predicts the change of state between any two times. That is done by numerically integrating a system of differential equations.

To simplify, it is convenient to write those differential equations in terms of alternative state variables from Sect.~\ref{sec:mono}. New values for the original state can be recovered after the integration. Specifically, we use the ``mean bulk'' $\bar W_s = B_s H_s / N_s$ and $\bar W_a = B_a H_a / N_a$, and the top heights $H_s$ and $H_a$ given by
\begin{subequations} \label{eq:hconv2}
\begin{align}
    H_s &=  \bar H_s \exp\left[0.10439 \max\left\{-\ln(150 \bar S_s), 0\right\}^{1.159}\right] \\ 
    H_a &=  \bar H_a \exp\left[0.04732 \max\left\{-\ln(150 \bar S_a), 0\right\}^{1.438}\right]
\end{align}
\end{subequations}
(c.~f.~eq.~\eqref{eq:hconv}). Some of the numerical coefficients derived from parameter estimates are given below at full precision to prevent rounding artifacts.

From eq.~\eqref{eq:spac}, the mean growing space $\bar S_s$ and $\bar S_a$ is
\begin{subequations} \label{eq:spac2}
\begin{align}
    \begin{split}
      \bar S_s &= \max \biggl\{\frac{1.317}{1.317 N_s + N_a} \Bigl[ 1 + 0.00002153 N_a (\bar H_s - \bar H_a)  \\
     &\qquad + 1.09469 \frac{N_s N_a}{(1.317 N_s + N_a)^2}\Bigr], 0\biggr\}
    \end{split} \\
    \begin{split}
      \bar S_a &= \max\biggl\{\frac{1}{1.317 N_s + N_a} \Bigl[ 1 + 0.00002153 N_s (\bar H_a - \bar H_s) \\
      & \qquad + 1.09469 \frac{N_s N_a}{(1.317 N_s + N_a)^2} \Bigr], 0\biggr\} \;.
    \end{split}
\end{align}
\end{subequations}
After integration, iterative numerical methods are needed to solve these relationships for the original variables.

Substituting parameter values in equations \eqref{eq:ode1} and  \eqref{eq:mix}, and using eq.~\eqref{eq:modif} and the aspen localization, the state transitions are given by the following system of differential equations:
\begin{subequations}  \label{eq:sam}
\begin{align}
  \begin{split}
    \der{H_s}{t} &= (q_s/0.5829) H_s \left[\left(283.87 q_s^{0.5137} / H_s\right)^{0.5829} - 1\right] \\
    & \qquad \left\{1 - \frac{0.4229 \left(N_a \max\{H_a - H_s, 0\}\right)^4}{(N_a^4 + 883667)[(H_a - H_s)^4 + 178.073]}\right\}	    
  \end{split} \\
  \begin{split}
    \der{H_a}{t} &= q_a H_a \left[(8.568 + 1342 q_a) / H_a - 1\right] \\
      & \qquad \left\{1 - \frac{0.4229 \left(N_s \max\{H_s - H_a, 0\}\right)^4}{(N_s^4 + 883667)[(H_s - H_a)^4 + 178.073]}\right\}
  \end{split} \\
  \der{\ln N_s}{t} &=  - \left(4.5759\times 10^{-15} H_s^{5.009} / \bar S_s^{1.9895}\right) \der{H_s}{t} \\
  \der{\ln N_a}{t} &=  - \left(0.000690992 / \bar S_a^{0.6159}\right) \der{H_a}{t} \\
  \der{\bar W_s}{t} &=  0.5345 \Omega H_s \bar S_s^{0.7} \der{H_s}{t} - 0.7 \bar W_s \der{\ln N_s}{t} \\
  \der{\bar W_a}{t} &=  1.29684 \Omega H_a \bar S_a^{0.7971} \der{H_a}{t} - 0.5 \bar W_a \der{\ln N_a}{t} 
\end{align}
\end{subequations}
with
\begin{subequations}  \label{eq:omega2}
\begin{align}
    \Omega_s &= 1 - (1 - \Omega_{0s}) \exp\left[-0.0258214 (H_s^2 - {H_{0s}}^2)\right] \\
    \Omega_a &= 1 - (1 - \Omega_{0a}) \exp\left[-0.015 (H_a^2 - {H_{0a}}^2)\right] \;,
\end{align}
\end{subequations}
where $H_{0s}, H_{0a}, \Omega_{0s}, \Omega_{0a}$ are values at the initial time.

The truncation at 0 in eq.~\eqref{eq:spac2} and in other functions produces a slope discontinuity that could cause numerical difficulties in optimization or zero-finding algorithms. Therefore, a smooth strictly positive hyperbolic approximation was substituted for the bound in those instances:
\begin{equation} \label{eq:positive}
  \max\{x, 0\} \approx \tfrac{1}{2} \left(x + \sqrt{x^2 + \epsilon^2}\right) \;.
\end{equation}  
The largest discrepancy is $\epsilon/2$ at $x=0$. For a good approximation this tolerance should be small enough to be practically negligible, but not so small as to generate numerical underflows. An $\epsilon$ close to the mean of $x$ times $10^{-4}$ seems adequate in most cases. A smoother transition obtained with larger values of $\epsilon$ might however be more realistic.

In addition to the R functions used in model fitting, SAM has been implemented as an easy-to-use interactive simulator. The simulator, using Microsoft Excel with Visual Basic macros, was based on those of \citet{scube,aspen} and \citet{loblolly}, and it is available in the Supplementary Material.

Avoiding numerical integration of the $\Omega$'s in the simulator was important, because these are fast variables that produce stiff differential equation systems, requiring sophisticated adaptive algorithms for accurate solutions \citep[e.~g.,][]{deSolve}. Making use of eq.~\eqref{eq:omega2}, a relatively simple modified 4th-order Runge-Kutta method with fixed time-step was sufficient. Careful debugging to achieve consistency of results among the simulator, the R functions, and the single-species models, makes serious programming errors unlikely. 

The SAM simulator produces tables of state variables, yields and other outputs for any sequence of times starting from a given state. By default, initial states at stand establishment are generated by the methods described in \ref{app:init}, including an optional spruce regeneration lag. Disturbances can be simulated at any time by introducing arbitrary changes in the state variables.

\ref{app:vol} derives total volume functions based on SAM's state variables. These, and merchantable volume relationships from Scube and TAG, are used to calculate stand volumes to any merchantability index given the current state variables. Other outputs like carbon or biomass can be easily added. Facilities are also provided to derive site quality from conventional site indices, optionally enforcing the species site equivalence of eq.~\eqref{eq:siteconv}.

    \subsection{Examples}
    \label{sec:ex}

Three examples are given to illustrate potential uses of the SAM simulator in some research and management problems, and to examine model behavior.

    \subsubsection{Replacement diagrams}
    \label{sec:diags}

Replacement diagrams or replacement series experiments were popularized by \citet{wit60} for studying plant competition or interference effects in two-species mixtures \citep[e.g.,][]{harper77,kelty92,jolliffe00,condes13,forrester15}. The diagram plots some measure of yield for each component and for the total of the mixture as functions of species composition, maintaining a constant total density. The original application of \citet{wit60} deals with seed yields at harvest in an annual crop, and both yields and composition are in terms of number of seeds per unit area. The situation is less clear in perennials \citep{wit65,firbank85}, and in forests in particular a variety of yield and species composition measures have been proposed \citep{dirnberger14,sterba14}. In addition to the density level, a harvest age has to be chosen.

For illustration purposes, an initial density of 5000 trees/ha (at breast height) and a harvest age of 100 years were specified. Spruce site index was 18, with the default 17.79 for aspen according to eq.~\eqref{eq:siteconv}. The default initialization procedure with spruce regeneration by planting was used (\ref{app:init}). Total volumes per hectare were computed with the SAM simulator for initial spruce numbers of 0, 10\%, 20\%, \dots, 100\%, but plotted over the more commonly used species composition \% by basal area. The result is shown in Fig.~\ref{fig:Figure5}. The customary straight lines drawn for comparison indicate the expected relationships in the absence of interactions, e.~g., on side-by-side monocultures.

\begin{figure}[htbp]
\fig{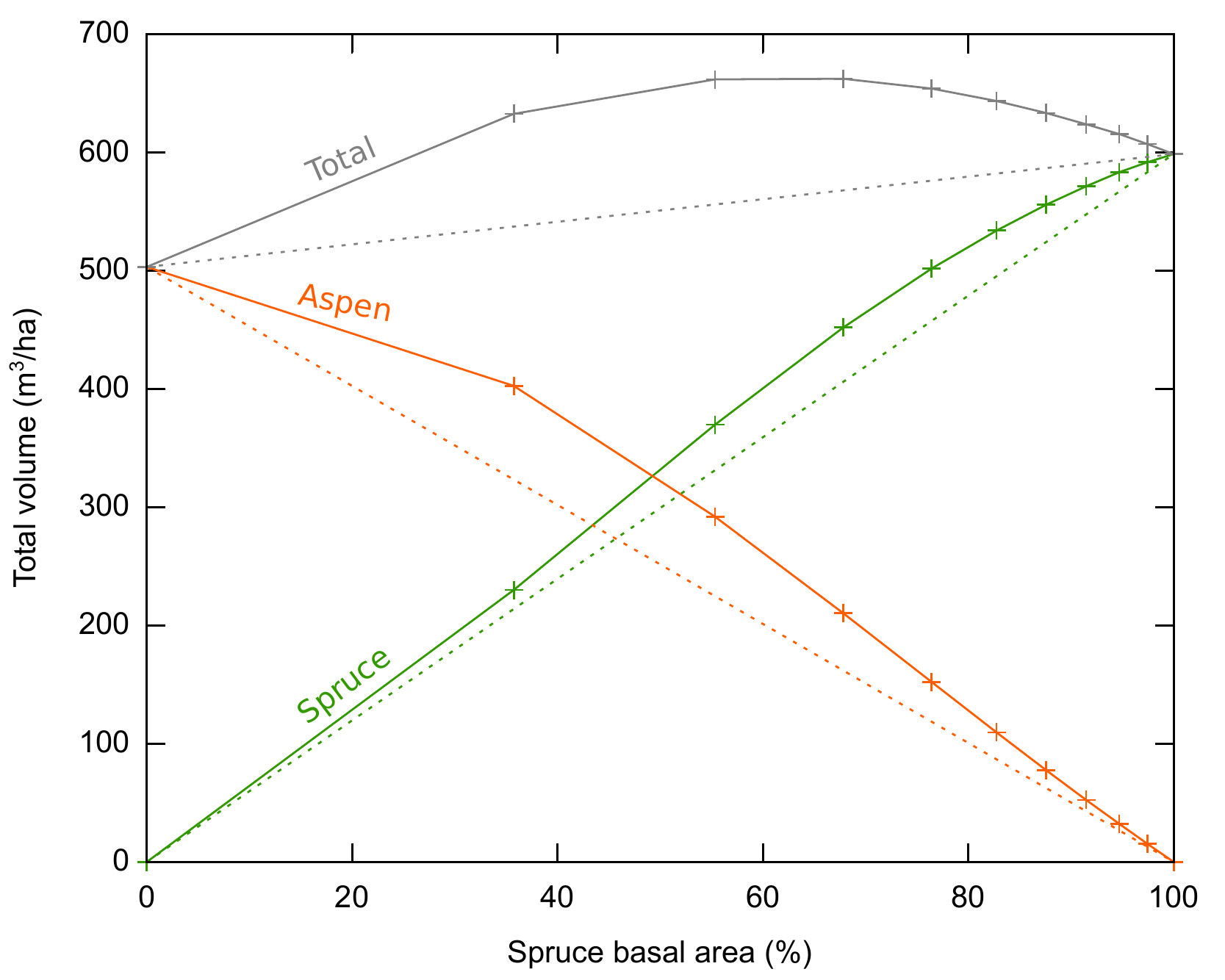}{Replacement diagram for age 100 years, 5000 trees/ha initial density, spruce site index 18. The dashed straight lines correspond to no interaction.} 
\end{figure}

The Figure is similar to typical replacement diagrams found in the literature. It exhibits higher yields in mixture than in monoculture, commonly termed \emph{overyielding}.

    \subsubsection{Optimization}
    \label{sec:opt}

As pointed out, among others, by \citet{firbank85}, \citet{jolliffe00}, and \citet{lilles14}, replacement series and other mixture experiments are often difficult to interpret because of confounding with stand density and other factors. The yield criterion may or may not represent management objectives and costs, and the best harvesting age may differ among the monocultures and the mixture. Moreover, basal area or other composition indicators are largely responses and not management decision variables. Growth models can provide more flexible means for addressing these issues \citep{firbank85}.

With SAM it is not difficult to choose management variables in a more realistic setting. The Excel \emph{Solver} facility can be used to compute their optimal values. As a simple example, it is shown how to maximize long-term CO2 sequestration in timber production. We consider only long term off-site carbon storage in forest products \citep{mckinley11}.

Merchantable volume per hectare to a merchantability limit of 15 cm was multiplied by basic density, 0.35 and 0.37 for spruce and aspen, respectively, to obtain wood production in metric tonnes per hectare \citep[][Table 4a]{WoodHandbook}. Carbon wood content is normally between 0.48 and 0.50, which can be multiplied by $44 / 12$ to convert to CO2 units. Dividing by the harvesting age gives a spreadsheet entry for mean annual CO2 sequestering in tons/ha--yr, which is selected in the Solver as the objective to be maximized.

Site indices and initialization were as in the previous example, except for the initial densities. The 3 decision variables in the Solver were the initial numbers of trees per hectare  for spruce and aspen, and the harvest age. Runs for monocultures were also performed, fixing the initial density at 0 for one of the two cohorts. Optimization takes a fraction of a second, and gave the results in Table \ref{tab:optim}.

\begin{table}[htbp]
  \tab[\footnotesize]{optim}{Maximizing carbon sequestration}{
  {lcccc}
    \hline
    Composition & CO2 (T/ha-yr) & Rotation (yr) & Initial spruce / ha & Initial aspen / ha \\
    \hline
    Mixture & 3.778 & 106.9 & 1573 & 6398  \\
    Spruce  & 3.729 & 94.1  & 892385  & 0  \\
    Aspen   & 2.872 & 83.5  & 0 & 1782 \\
    \hline
    }
\end{table}

Optimization is a stringent test, being ``remarkably efficient at exploiting seemingly minor quirks in models to arrive at unrealistic solutions''
\citep{vanclay97}. That was the case for initial density in the spruce monoculture.

    \subsubsection{Spruce regeneration lags}
    \label{sec:lags}
    
Delays in the regeneration of a cohort can have important consequences for stand dynamics. In particular, spruce regeneration lags after a stand-replacing fire can produce distinct dominance patterns, which can be mistaken by chronosequence stages in the so-called classical dynamics successional pathway \citep{kabzems04,johnson08,bergeron14}. The SAM simulator implements a spruce regeneration delay that can be used to study these issues.

\begin{figure}[htbp]
  \fig{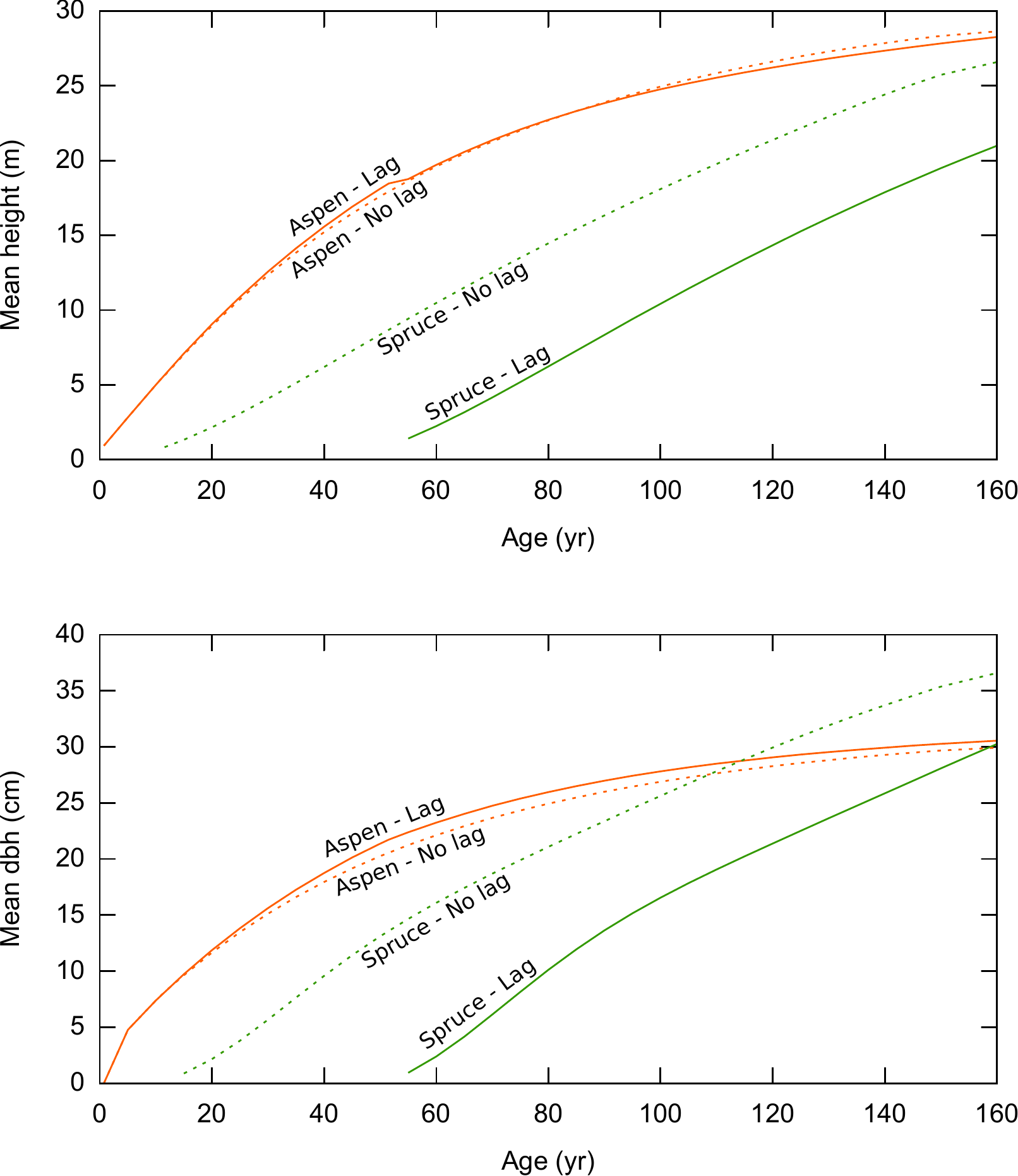}{Simulated mean height and mean diameter trends with and without a spruce regeneration lag (c.~f.\ \citet{kabzems04}).} 
\end{figure}

Figure \ref{fig:Figure6} shows simulated mean heights and diameters with a 40-yr spruce regeneration lag, compared to projections without a lag. Site indices were 19 and 21, and initial densities 1000 and 8000 trees / ha, for spruce and aspen, respectively. Mean height and diameter trends are qualitatively similar to those reconstructed through stem analysis by \citet{kabzems04}, although the lag effects appear somewhat smaller than suggested by that study.

\section{Discussion and Conclusions}
\label{sec:disc}

The process of data preparation and preliminary analysis pointed to some interesting patterns. The breakup often reported in mature aspen stands \citep{frey04} was not evident in our data from British Columbia. Aspen diameters and basal area for any given height tended to be higher than those predicted by the TAG model for the whole of western Canada, an observation confirmed by a statistically significant localization parameter. \citet{kabzems04} commented on the large size of individual trees, longevity, and the low occurrence of internal decay in trembling aspen in northeastern British Columbia compared to other boreal forests. To some extent this may apply to the rest of the Province, although biogeoclimatic zone comparisons with the available data are inconclusive.

On the other hand, there were clear indications of heavy mortality among older spruce trees. Such spruce old-growth stand breakup does not seem to have been previously reported in the literature. If confirmed, it can have important implications for sustainable forest management and wood supply, so that further research on this matter is desirable.

It is worth mentioning also the effectiveness of the ingrowth imputation procedure. Similar methods can avoid awkward ingrowth modelling problems, and make better use of young stand data whenever trees are not measured below some size threshold, a common practice in natural forests.

Current individual-based methods present problems with emergent properties and statistical confounding that have not yet been generally recognized. These issues deserve attention, given the undeniable value of those methods for studying the ecological fundamentals of stand dynamics. Meanwhile, aggregation has the potential for side-stepping some of those difficulties and providing adequate management tools, even if only as a stop-gap measure. However, although aggregation reduces the number of variables, it increases the level of abstraction, making less obvious the mapping between modelling elements and the real world compared to the typical individual-tree model. That has hampered the development of mechanistic aggregate stand-level models for other that single-species even-aged stands. This work is a preliminary exploration of a conceptual framework and methods for modelling complex stands at the stand level.

Turning to the model components, already available single-species models served as a starting point, linking the cohort models through space partitioning relationships. In other situations, the single-species models would need to be developed from scratch, possibly helped by the perfect plasticity theory from section \ref{sec:size}, as outlined on p.~11 of \citet{plasticity}. In any case, when there is a need for predictions beyond the environmental and management conditions represented in the available data, one should avoid the common practice of using stem thickness measures such as diameter or basal area as driving variables \citep{loblolly,lee16}

Resource capture or growing space is a highly simplified abstraction of ecological processes, but it seems a useful basis for modelling and understanding interactions in forest stands \citep{monteith94}. The space partitioning model of \citet{wit60} was extended to deal with distinct tree sizes, and to allow for differences in resource use efficiency between mixtures and monocultures (complementarity). In the spruce-aspen stands, de Wit's relative crowding coefficient of spruce with respect to aspen was estimated as $k = 1.317$, with a statistically significant difference from 1, indicating that spruce trees capture more site resources than equally tall aspens. The term accounting for the effect of tree height differences was also significant, with substantially more space allocated to taller cohorts. There was support for enhanced efficiency in spruce-aspen mixtures, with a statistically significant estimated increase of up to 21\% in resource availability relative to monocultures. The space allocation model makes possible an objective testing of theoretical arguments and assumptions common in the literature.

Mean cohort heights, suitable as mixture state variables, could be related satisfactorily to stand density and the more convenient top heights used in single-species even-aged models. Relationships similar to those developed in \ref{app:topmean} would be useful in other forestry applications. Including height growth suppression complicated substantially the model structure and parameter estimation, but it clearly improved predictions in the more highly stratified plots. It was necessary for approximating the observed behavior under regeneration lags described in Sec.~\ref{sec:lags}. 

Parameter estimation in this model is unusually complex. It involves multiresponse, with panel data, in a hierarchical model constituted by a system of nonlinear differential equations that need to be numerically integrated. Each evaluation requires also iterative root finding. With common distributional assumptions, minimizing the Box-Draper multiresponse determinant criterion corresponds to maximum likelihood or Bayesian estimation, and it has the good asymptotic properties of generalized least-squares under less restrictive conditions. It is an interesting reflection on modern computing power that estimates could be obtained with standard numerical analysis software on a personal computer, in a reasonable amount of time and without major difficulties. Satisfactory fits were verified by analysis of residuals, parameter statistics and simulations 

The spruce-aspen simulator demonstrates a user-friendly and flexible implementation. Most users are already familiar with spreadsheet software, and the complexity of the calculations is hidden in efficiently coded  macros that execute quickly in the background. Besides showing the model's plausible behavior, the examples of Sect.~\ref{sec:ex} suggest the nature of possible applications. In particular, built-in optimization tools can easily answer questions that until recently would have been regarded as sizable research projects.

SAM, the spruce-aspen model, must be viewed largely as an experiment for testing the methodology. Still, it can be useful for management planning and for guiding research if its limitations are kept in mind. Suitable data from mixed stands was rather scarce, and the model would benefit from further testing with new experimental data that is gradually becoming available \citep[e.~g.,][]{bokalo07,gradowski08,comeau09,kabzems15}. SAM is not expected to be reliable in old-growth stands due to the unpredictability of die-back in mature aspen \citep{frey04,hogg08,worrall13}, and possibly in older spruce (Sec.~\ref{sec:data}). Reliability is also uncertain in early stand development, where competition or facilitation mechanisms and relationships may differ from those in older stands, and where the effects of understory vegetation can be important. In some instances, a third cohort for the herbaceous vegetation might better reflect the interactions \citep{cosmin08}. The general approach, however, seems useful as a basis for future work with these and other species or stand types.

\section*{Acknowledgements} \addcontentsline{toc}{section}{Acknowledgements}
\label{sec:ack}

The data were provided by the Forest Analysis and Inventory Branch of the B.~C.~Ministry of Forests, Lands and Natural Resource Operations, and I am particularly grateful to Rene De Jong and Pat Martin for help in data extraction and interpretation. Thanks also to the Virginia Tech Forest Modeling Research Cooperative for permission to use the loblolly pine data in \ref{app:topmean}. Comments from anonymous referees contributed to improve the text. 

\appendix

\section{Growing space partitioning}
\label{app:part}

This appendix extends the relationships of Section \ref{sec:size} to any number of cohorts and to other than parabolic profiles.

Let the growing space or resource captured by tree $i$ in cohort $j$ be
\begin{equation} \label{eq:treeS}
  S_{ij} = b_j (H_{ij} - z^*)^\theta  \;,
\end{equation}
where $H_{ij}$ is the tree height, $z^*$ is the closure level, and $b_j$ and $\theta$ are positive parameters (Fig.~\ref{fig:Figure3}). This assumes that $H_{ij} \ge z^*$, otherwise the tree is completely over-topped and does not receive any resources. Any such trees are excluded from the calculations that follow, which might require some iterations since $z^*$ is not known in advance.

The cohort means are
\begin{equation} \label{eq:cohS}
  \bar S_j = b_j \overline{(H_{ij} - z^*)^\theta} \;,
\end{equation}
with the overline indicating a mean over the trees in cohort $j$. The level $z^*$ can be obtained from the fact that, with full closure (i.~e., $z^* \ge 0$), the individual values add up to the total available resource $R$:
\[
  \sum_{ij} S_{ij} = \sum_j N_j \bar S_j = R \;.
\]
In what follows, the number of trees $N_j$ and $R$ are quantities per unit area.
Therefore,
\begin{equation} \label{eq:sumS}
  \sum_j N_j b_j \overline{(H_{ij} - z^*)^\theta} = R \;.
\end{equation}
This equation can be solved for $z^*$, generally by numerical methods.

If $z^*$ is found to be negative, it can be taken as defining a potential resource capture, using then other means for modelling stand openness, as in SAM. Alternatively, $z^*$ can be bound at 0, giving a less than full site occupancy in eq.~\eqref{eq:cohS}:
\begin{equation} \label{eq:open}
  \bar S_j = b_j \overline{H_{ij}^\theta} \;.
\end{equation}
The approximation in eq.~\eqref{eq:positive} could be used to preserve smoothness.

Explicit results can be obtained for $\theta = 1$ and for $\theta = 2$.

\subsection*{Case $\theta = 1$}
\label{sec:theta1}

Eq.~\eqref{eq:cohS} becomes
\begin{equation} \label{eq:cohS2}
  \bar S_j = b_j (\bar H_j - z^*) \;,
\end{equation}
and from eq.~\eqref{eq:sumS},
\[
    \sum_j N_j b_j \bar H_j - z^* \sum_j N_j b_j = R \;,
\]
so that
\begin{equation} \label{eq:z1}
  z^* = \frac{\sum_j N_j b_j \bar H_j - R}{\sum_j N_j b_j} \;.
\end{equation}

\subsection*{Case $\theta = 2$}
\label{sec:theta2}

Now eq.~\eqref{eq:cohS} becomes
\begin{equation} \label{eq:cohS3}
  \bar S_j = b_j (\bar H_j - z^*)^2 \;,
\end{equation}
and eq.~\eqref{eq:sumS} is
\[
    \sum_j N_j b_j \overline{(H_{ij} - z^*)^2} = R \;.
\]
This gives a quadratic equation in $z^*$
\begin{equation} \label{eq:quad}
  (z^*)^2 \sum_j N_j b_j - 2 z^* \sum_j N_j b_j \bar H_j +
    \sum_j N_j b_j \overline{H_{ij}^2} - R = 0
\end{equation}
that can be analytically solved for $z^*$.

For a somewhat simpler expression, define $\tilde H$ as the weighted mean
\begin{equation} \label{eq:tildeH}
      \tilde H = \frac{\sum_j N_j b_j \bar H_j}{\sum_j N_j b_j} \;,
\end{equation}
and
\begin{equation} \label{eq:sigma}
  \tilde\sigma^2 = \frac{\sum_j N_j b_j \tilde\sigma^2_j}{\sum_j N_j b_j} = \frac{\sum_j N_j b_j \overline{(H_{ij} - \tilde H)^2}}{\sum_j N_j b_j} \;,
\end{equation}
a weighted variance.
Then write
\begin{align*}
      R &= \sum_j N_j b_j \overline{(H_{ij} - z^*)^2}
      = \sum_j N_j b_j \overline{[(H_{ij} - \tilde H) + (\tilde H - z^*)]^2}
      \\ &= \sum_j N_j b_j [\overline{(H_{ij} - \tilde H)^2} + (\tilde H - z^*)^2]
      \\ &= [\tilde\sigma^2 + (\tilde H - z^*)^2] \sum_j N_j b_j
\end{align*}
(the cross-products term vanishes).
Finally, solving for $z^*$,
\begin{equation} \label{eq:z2}
  z^* = \tilde H - \sqrt{\frac{R}{\sum_j N_j b_j} - \tilde\sigma^2} \;.
\end{equation}

\section{Top height \emph{vs.} mean height}
\label{app:topmean}

Measurements with at least 95\% of basal area in either spruce or aspen were selected to obtain relationships between top height $H$, mean height $\bar H$, and trees per hectare $N$. Given the small size of these samples, a much larger data set from loblolly pine plantations used by \citet{loblolly} helped in testing for appropriate equation forms.

Data statistics are shown in Table \ref{tab:hdata}. Another interesting characteristic is the height variability within plots. The mean coefficient of variation was 22.3\% for spruce, 15.8\% for aspen, and 9.9\% for pine. Heights were estimated from height-diameter regressions in spruce and aspen, but measured directly on all trees in the pine data.

\begin{table}[htbp]
  \tab{hdata}{Data for height relationships. Trees per hectare $N$, and mean height $\bar H$ (m).}{
  {lcccccc}
      \hline
      & \multicolumn{2}{c}{Spruce ($n=26$)} & \multicolumn{2}{c}{Aspen ($n=52$)} & \multicolumn{2}{c}{Pine ($n=1156$)} \\
      & $N$ & $\bar H$ & $N$ & $\bar H$ & $N$ & $\bar H$ \\
      \hline
      Mean  & 1380  & 13.34 & 2481  & 17.34 & 951   & 15.37 \\
      S.~D. & 536   & 7.67  & 1407  & 4.75  & 430   & 3.74  \\
      Min.  & 568   & 3.34  & 338   & 10.67 & 253   & 5.75  \\
      Max.  & 2275  & 30.94 & 7075  & 29.50 & 2644  & 26.07 \\
      \hline
  }
\end{table}

Equations were conditioned to give a top height equal to the mean height at 150 trees / ha, the approximate top trees selection rate $160 - 0.6/A$ for our mean plot size $A = 0.09$ ha \citep{topht}.  The logarithm of the height ratio $\log(H / \bar H)$ was used as dependent variable in the regressions, because it gives the same parameter estimates when estimating $H$ from $\bar H$ or vice-versa, and because of improved homocedasticity.

Equation forms were screened by comparison to theoretical relationships generated from Weibull distributions, and to the empirical data. Best results were obtained with  $\ln(H / \bar H) = a (N - 150)^b$, $\ln(H / \bar H) = a [\ln (N / 150)]^b$, $\ln(H / \bar H) = a [(N / 150)^b - 1]$, and with $\ln(H / \bar H) = a [\sqrt{N / 150} - 1]$, all these equations producing very similar fit statistics. The second form was finally chosen, written as
\begin{equation} \label{eq:topmean}
  \ln(H / \bar H) = a \max\{\ln (N / 150), 0\}^b
\end{equation}
to include values of $N < 150$. The nonlinear least squares estimates are given in Table \ref{tab:hreg}. For mixtures, the reciprocal $1 / \bar S$ of the mean space per tree was substituted for $N$ (eq.~\eqref{eq:hconv}).

\begin{table}[htbp]
  \tab{hreg}{Nonlinear regressions for eq.~\eqref{eq:topmean}}{
    {lcccc}
        \hline
        Species & $n$ & $\hat a$ & $\hat b$ & RSE \\
        \hline
        Spruce  & 26    & 0.10439 & 1.159 & 0.07265 \\
        Aspen   & 52    & 0.04732 & 1.438 & 0.07685 \\
        Pine    & 1156  & 0.03466 & 1.487 & 0.02746 \\
        \hline 
    }
\end{table}

\section{Simulation initialization}
\label{app:init}

SAM should be most reliable projecting the development of closed-canopy stands starting from a known state. However, it is also interesting to simulate stand dynamics from stand initiation. In the single-species models it was convenient to initialize at breast-height age, but the fact that spruce reaches breast height later than aspen complicates modelling for mixed stands. The effect of lags in spruce regeneration is also important \citep{kabzems04,bergeron14}. The implemented simulator provides the following initialization procedure by default, although other options can be used.

An aspen stand is started with a top height of 1.3 m (breast height), a user-supplied number of trees, basal area 0, and occupancy estimated as a function of the number of trees as in Section 3.3.3 of \citet{aspen}. The time origin is set at stand initiation, with the time at breast height being
\begin{equation} \label{eq:ytbh-aspen}
    -\frac{1}{q} \ln\left(1 - \frac{1.3}{8.568 + 1342 q}\right) - 1.183
\end{equation}
\citep[][eq.~5]{aspen}.

Spruce is similarly introduced when it reaches breast height, at time $7.7 + 111/\text{\it site index}$ for natural seed-origin stands, or $1.7 + 111/\text{\it site index}$ if the spruce is planted \citep{hu10}. A spruce regeneration lag can be added if regeneration does not occur at the time of stand initiation. In this case growing space and mean heights are calculated for the mixture with equations \eqref{eq:spac} and \eqref{eq:hconv}, and the reciprocal of the spruce growing space is used instead of number of trees in the occupancy equations of \citet{scube}. This implies neglecting any effects of spruce on aspen growth before the spruce reaches breast height.

As noted elsewhere, SAM might not represent well seedling development below breast height. Competition, facilitation and complementarity mechanisms can differ among life stages, the empirical single-species occupancy relationships may not be accurate for mixtures, and the presence of secondary vegetation and soil and vegetation control treatments can be important factors. Simulations from stand initiation can be useful, but currently their accuracy is uncertain.

\section{Volume equations}
\label{app:vol}

The total volume per hectare $V$ for each species needs to be estimated from the state variables. Table \ref{tab:vdata} gives statistics for the measurements available. All measurements containing the target species were used, regardless of mixing proportions. 

\begin{table}[htbp]
  \tab{vdata}{Data for volume regressions. Mean height $\bar H$ (m), basal area $B$ (m$^2$ / ha), and total volume $V$  (m$^3$ / ha).}{
  {lcccccc}
      \hline
      & \multicolumn{3}{c}{Spruce ($n=401$)} & \multicolumn{3}{c}{Aspen ($n=389$)} \\
      & $\bar H$ & $B$ & $V$  & $\bar H$ & $B$ & $V$  \\
      \hline
      Mean  & 16.70  & 23.88 & 224.6  & 18.69  & 17.59 & 146.7 \\
      S.~D. & 7.49   & 15.81  & 187.0  & 5.25 & 13.66 & 128.4  \\
      Min.  & 3.34   & 0.02  & 0.0   & 6.30 & 0.07  & 0.2  \\
      Max.  & 35.05  & 70.95 & 904.3  & 35.13 & 63.28 & 789.6 \\
      \hline
  }
\end{table}
 
A large number of regressions and weightings were tested through stepwise regression and graphical analysis. Species composition had no appreciable effect, and neither had the inclusion of number of trees or growing space. That left the cohort basal area $B$ and the mean height $\bar H$ as suitable predictors. Using the form factor $V / (B\bar H)$ as dependent variable gave good results, and little improvement was obtained with relationships more complicated than simple linear regressions on $\bar H$. The final models, with the subscripts $s$ and $a$ for spruce and aspen, respectively, were:
\begin{align}
  \frac{V_s}{B_s \bar H_s} &= 0.5873 - 0.005386 \bar H_s \;, \quad \text{RSE: } 0.07261   \label{eq:vs} \\
  \frac{V_a}{B_a \bar H_a} &= 0.4765 - 0.002196 \bar H_a \;, \quad \text{RSE: } 0.02894   \label{eq:va}
\end{align}
In the simulator, negative values are suppressed as $V / (B \bar H) = \max\{b_0 - b_1 \bar H, 0\}$, smoothing the transition with eq.~\eqref{eq:positive}.

\bibliographystyle{elsarticle-harv}
\bibliography{sam} \addcontentsline{toc}{section}{References}

\end{document}